\documentclass[12pt]{article}

\usepackage{cite}
\usepackage{graphicx}
\usepackage{pict2e}

\author{Yu.~M.~Zinoviev
       \thanks{E-mail address: Yurii.Zinoviev@ihep.ru} \\[0.5cm]
        {\it Institute for High Energy Physics} \\
        {\it of National Research Center "Kurchatov Institute"} \\
        {\it Protvino, Moscow Region, 142280, Russia}}

\title{On massive higher spins and gravity. II. Spin 3}

\date{}

\begin{document}

\maketitle

\begin{abstract}
In this paper, we continue our investigation of gravitational
interactions for massive higher spins, extending our recent work on
massive spin 5/2 to massive spin 3, including its massless and
partially massless limits. To construct the minimal gravitational
interactions (i.e. vertexes containing both standard minimal
interactions and non-minimal ones, which are necessary for
any $s \ge 5/2$), we use a gauge invariant frame-like formalism.
Similarly to the spin 5/2 case, there is a special point 
$m^2 = 6\Lambda$, which corresponds to a boundary of the
unitary allowed region in $dS_4$, where minimal interactions 
disappear, leaving only the non-minimal ones.
\end{abstract}

\thispagestyle{empty}
\newpage
\setcounter{page}{1}

\section{Introduction}

Recently, in \cite{Zin25a} we investigated the gravitational
interactions for massive spin 5/2 (the first case that requires
non-minimal corrections), including its massless and partially
massless limits. To maintain the correct number of physical degrees of
freedom, we used gauge invariant frame-like formalism
\cite{Zin08b,PV10,KhZ19} for to describe free massive spin 5/2 and the
so-called  Fradkin-Vasiliev formalism
\cite{FV87,FV87a,Vas11,KhZ20,Zin24a} to construct interactions. Using
the gauge invariant formalism for massive fields, we encountered
ambiguities related with the field redefinitions containing
Stueckelberg fields. In particular, any vertex for one massless field
and two massive ones (as is the case for the gravitational
interactions) can be converted into a purely abelian form. To solve
these ambiguities, and construct a minimal vertex (i.e. containing
standard minimal interactions and the non-minimal ones with
the minimum number of derivatives) we used a common down-up
approach. It turned out to be the most efficient way to determine the
general structure of the non-minimal interactions y considering the
partially massless case in the Skvortsov-Vasiliev version
\cite{SV06,KhZ19} of the frame-like formalism. Using these findings,
we managed to construct a minimal gravitational vertex in the general
massive case. Our results have a non-singular massless limit for a
non-zero cosmological constant and a non-singular flat limit with
non-zero mass. There is a specific point that corresponds to
the  boundary of the unitary allowed region in $dS_4$, where standard
minimal interactions are absent and only non-minimal remain. Our
goal here is to apply the same approach for massive spin 3 (the
first bosonic case that needs non-minimal corrections), including
its massless and partially massless limits.

The paper is organized as follows. In Section 2 we provide all
necessary kinematical information about the gauge invariant frame-like
description of massive spin 3 and its massless and partially massless
limits. In Section 3, we construct a gravitational vertex for the
massless case,  using the Fradkin-Vasiliev formalism. In Section 4, we
proceed with the partially massless case by using a Skvortsov-Vasiliev
version of the frame-like formalism\footnote{There are no ambiguities
related with the field redefinitions in this version, so the
Fradkin-Vasiliev formalism works exactly as in the massless case.}.
This gives us with the general structure for non-minimal interactions,
and allows us in Section 5 to construct a minimal gravitational vertex
for the massive spin 3. 

\section{Kinematics}

In this Section, we provide all the necessary kinematic information on
massive spin 3, including its massless and partially massless limits.
We work in a gauge invariant frame-like multispinor formalism
\cite{KhZ19} and use the same notation and conventions as in
\cite{Zin25a}.

\subsection{Massive spin 3}

A massive spin 3 has helicities $(\pm 3, \pm 2, \pm 1, 0)$ therefore a
gauge invariant frame-like description requires four pairs
(auxiliary field, physical field): 
$(\Omega^{\alpha(3)\dot\alpha} + h.c., f^{\alpha(2)\dot\alpha(2)})$,
$(\Omega^{\alpha(2)} + h.c., f^{\alpha\dot\alpha})$,
$(B^{\alpha(2)} + h.c., A)$ and $(\pi^{\alpha\dot\alpha}, \varphi)$.
As is always the case for the gauge invariant description of massive
bosons, the free Lagrangian consists of three parts: kinetic terms,
cross terms and mass terms:
\begin{equation}
{\cal L}_0 = {\cal L}_{kin} + {\cal L}_{cros} + {\cal L}_{mass},
\end{equation}
\begin{eqnarray}
{\cal L}_{kin} &=& - 3 \Omega^{\alpha(2)\beta\dot\alpha}
E_\beta{}^\gamma \Omega_{\alpha(2)\gamma\dot\alpha} + 
\Omega^{\alpha(3)\dot\alpha} E_{\dot\alpha}{}^{\dot\beta}
\Omega^{\alpha(3)\dot\beta} - 2 \Omega^{\alpha(2)\beta\dot\alpha}
e_\beta{}^{\dot\beta} D f_{\alpha(2)\dot\alpha\dot\beta} \nonumber \\
 && + 2 \Omega^{\alpha\beta} E_\beta{}^\gamma \Omega_{\alpha\gamma}
+ 2 \Omega^{\alpha\beta} e_\beta{}^{\dot\alpha} D f_{\alpha\dot\alpha}
 + 4E B_{\alpha(2)} B^{\alpha(2)} + 2 E_{\alpha(2)} B^{\alpha(2)}
D A \nonumber \\
 && - 3 E \pi_{\alpha\dot\alpha} \pi^{\alpha\dot\alpha} - 6
E_{\alpha\dot\alpha} \pi^{\alpha\dot\alpha} D \varphi + h.c.,
\label{lag_1} \\
{\cal L}_{cros} &=& 4\tilde{m} [\Omega^{\alpha\beta(2)\dot\alpha}
E_{\beta(2)} f_{\alpha\dot\alpha} + \frac{1}{3} 
f^{\alpha(2)\dot\alpha(2)} E_{\beta(2)} \Omega_{\dot\alpha(2)}]
\nonumber \\
 && + a_0 [ \Omega^{\alpha(2)} E_{\alpha(2)} A - 2 B^{\alpha\beta}
E_\beta{}^{\dot\alpha} f_{\alpha\dot\alpha} ] + \tilde{a}_0
E_{\alpha\dot\alpha} \pi^{\alpha\dot\alpha}A + h.c., \label{lag_2} \\
{\cal L}_{mass} &=& - M^2 f^{\alpha\beta\dot\alpha(2)} 
E_\beta{}^\gamma f_{\alpha\gamma\dot\alpha(2)} + 4M^2
f^{\alpha\dot\alpha} E_\alpha{}^\beta f_{\beta\dot\alpha} + h.c.
\nonumber \\
 && + \frac{a_0\tilde{a}_0}{2} E_{\alpha\dot\alpha}
f^{\alpha\dot\alpha} \varphi + 3a_0{}^2 E \varphi^2. \label{lag_3}
\end{eqnarray}
Here
\begin{equation}
M^2 = m^2 - 6 \Lambda, \qquad
\tilde{m}^2 = \frac{3m^2}{4}, \qquad
a_0{}^2 = 20[m^2 - 4\Lambda], \qquad
\tilde{a}_0{}^2 = 288M^2. \label{unitary}
\end{equation}
This Lagrangian is invariant under the following gauge transformations
for the physical fields:
\begin{eqnarray}
\delta f^{\alpha(2)\dot\alpha(2)} &=& D \xi^{\alpha(2)\dot\alpha(2)}
 + e_\beta{}^{\dot\alpha} \eta^{\alpha(2)\beta\dot\alpha}
+  e^\alpha{}_{\dot\beta} \eta^{\alpha\dot\alpha(2)\dot\beta} 
+ \frac{\tilde{m}}{3} e^{\alpha\dot\alpha} \xi^{\alpha\dot\alpha},
\nonumber \\
\delta f^{\alpha\dot\alpha} &=& D \xi^{\alpha\dot\alpha} +
e_\beta{}^{\dot\alpha} \eta^{\alpha\beta} + e^\alpha{}_{\dot\beta}
\eta^{\dot\alpha\dot\beta} + \frac{2\tilde{m}}{3} e_{\beta\dot\beta}
\xi^{\alpha\beta\dot\alpha\dot\beta} - \frac{a_0}{4}
e^{\alpha\dot\alpha} \xi, \label{gaug_1} \\
\delta A &=& D \xi - \frac{a_0}{2} e_{\alpha\dot\alpha}
\xi^{\alpha\dot\alpha}, \qquad 
\delta \varphi = \frac{\tilde{a}_0}{12}, \xi \nonumber
\end{eqnarray}
and for the auxiliary fields:
\begin{eqnarray}
\delta \Omega^{\alpha(3)\dot\alpha} &=& D \eta^{\alpha(3)\dot\alpha}
+ e_\beta{}^{\dot\alpha} \eta^{\alpha(3)\beta} 
 + \frac{\tilde{m}}{6}e^{\alpha\dot\alpha} \eta^{\alpha(2)} +
\frac{M^2}{6} e^\alpha{}_{\dot\beta}
\xi^{\alpha(2)\dot\alpha\dot\beta}, \nonumber \\
\delta \Omega^{\alpha(2)} &=& D \eta^{\alpha(2)} + 2\tilde{m}
e_{\beta\dot\alpha} \eta^{\alpha(2)\beta\dot\alpha} + M^2
e^\alpha{}_{\dot\alpha} \xi^{\alpha\dot\alpha}, \label{gaug_2} \\
\delta B^{\alpha(2)} &=& \frac{a_0}{2}\eta^{\alpha(2)}, \qquad
\delta \pi^{\alpha\dot\alpha} = -
\frac{a_0\tilde{a}_0}{24}\xi^{\alpha\dot\alpha}. \nonumber
\end{eqnarray}
But to construct a complete set of the gauge invariant curvatures
(two forms for gauge one-forms and one-forms for Stueckelberg
zero-forms) we need so-called extra fields (which do not appear in the
free Lagrangian) with the following gauge transformations:
\begin{eqnarray}
\delta \Omega^{\alpha(4)} &=& D \eta^{\alpha(4)} + \frac{a_0{}^2}{80} 
e^\alpha{}_{\dot\alpha} \eta^{\alpha(3)\dot\alpha}, \qquad
\delta B^{\alpha(4)} = \eta^{\alpha(4)}, \nonumber \\
\delta B^{\alpha(3)\dot\alpha} &=& \eta^{\alpha(3)\dot\alpha}, \qquad
\delta \pi^{\alpha(2)\dot\alpha(2)} = \xi^{\alpha(2)\dot\alpha(2)}.
\end{eqnarray}
Thus a complete set of fields can be represented as follows:
$$
\begin{array}{|c|c|c|c|} \hline
one-forms & f^{\alpha(2)\dot\alpha(2)} & f^{\alpha\dot\alpha} & A \\
\hline zero-forms & \pi^{\alpha(2)\dot\alpha(2)} &
\pi^{\alpha\dot\alpha} & \varphi \\ \hline \end{array}  
\qquad
\begin{array}{|c|c|c|c|} \hline
one-forms & \Omega^{\alpha(4)} & \Omega^{\alpha(3)\dot\alpha} & 
\Omega^{\alpha(2)} \\ \hline zero-forms & B^{\alpha(4)} &
B^{\alpha(3)\dot\alpha} & B^{\alpha(2)} \\ \hline \end{array} + h.c.
$$
We see that there is a one-to-one correspondence between one-forms and
zero-forms, which is natural because in the massive case all gauge
symmetries must be spontaneously broken and each gauge one-form must
has its own Stueckelberg zero-form. 

Having explicit forms for all the gauge transformations, it is a
straightforward task to construct the complete set of the gauge
invariant curvatures. For the physical fields we obtain
\begin{eqnarray}
{\cal T}^{\alpha(2)\dot\alpha(2)} &=& D f^{\alpha(2)\dot\alpha(2)}
+ e_\beta{}^{\dot\alpha} \Omega^{\alpha(2)\beta\dot\alpha}
+ e^\alpha{}_{\dot\beta} \Omega^{\alpha\dot\alpha(2)\dot\beta}
+ \frac{\tilde{m}}{3} e^{\alpha\dot\alpha} f^{\alpha\dot\alpha},
\nonumber \\
{\cal T}^{\alpha\dot\alpha} &=& D f^{\alpha\dot\alpha} 
+ e_\beta{}^{\dot\alpha} \Omega^{\alpha\beta} + e^\alpha{}_{\dot\beta}
\Omega^{\dot\alpha\dot\beta} + \frac{2\tilde{m}}{3} e_{\beta\dot\beta}
f^{\alpha\beta\dot\alpha\dot\beta} - \frac{a_0}{4}
e^{\alpha\dot\alpha} A, \nonumber \\
{\cal A} &=& D A + 2 (E_{\alpha(2)} B^{\alpha(2)} + E_{\dot\alpha(2)}
B^{\dot\alpha(2)}) - \frac{a_0}{2} e_{\alpha\dot\alpha}
f^{\alpha\dot\alpha}, \label{cur_1} \\
{\cal C} &=& D \varphi + e_{\alpha\dot\alpha} \pi^{\alpha\dot\alpha}
- \frac{\tilde{a}_0}{12} A, \nonumber
\end{eqnarray}
while for the auxiliary fields we get
\begin{eqnarray}
{\cal R}^{\alpha(3)\dot\alpha} &=& D \Omega^{\alpha(3)\dot\alpha}
 + e_\beta{}^{\dot\alpha} \Omega^{\alpha(3)\beta} 
+ \frac{\tilde{m}}{6}e^{\alpha\dot\alpha} \Omega^{\alpha(2)} +
\frac{M^2}{6} e^\alpha{}_{\dot\beta} f^{\alpha(2)\dot\alpha\dot\beta},
\nonumber \\
{\cal R}^{\alpha(2)} &=& D \Omega^{\alpha(2)} + 2\tilde{m}
e_{\beta\dot\beta} \Omega^{\alpha(2)\beta\dot\beta} + M^2
e^\alpha{}_{\dot\alpha} f^{\alpha\dot\alpha} - \frac{a_0}{4}
E^\alpha{}_\beta B^{\alpha\beta} + \frac{a_0\tilde{a}_0}{24}
E^{\alpha(2)} \varphi \nonumber \\
 && - 3\tilde{m} E_{\beta(2)} B^{\alpha(2)\beta(2)} 
+ \frac{8M^2\tilde{m}}{3} E_{\dot\alpha(2)} 
\pi^{\alpha(2)\dot\alpha(2)}, \label{cur_2} \\
{\cal B}^{\alpha(2)} &=& D B^{\alpha(2)} - \frac{a_0}{2}
\Omega^{\alpha(2)}- \frac{\tilde{a}_0}{24} e^\alpha{}_{\dot\alpha}
\pi^{\alpha\dot\alpha} + \frac{a_0a_3}{4} e_{\beta\dot\alpha}
B^{\alpha(2)\beta\dot\alpha}, \nonumber \\
\Pi^{\alpha\dot\alpha} &=& D \pi^{\alpha\dot\alpha} - 
\frac{\tilde{a}_0}{12} (e_\beta{}^{\dot\alpha} B^{\alpha\beta} +
e^\alpha{}_{\dot\beta} B^{\dot\alpha\dot\beta}) 
+ \frac{a_0\tilde{a}_0}{24} f^{\alpha\dot\alpha} + \frac{a_0{}^2}{8}
e^{\alpha\dot\alpha} \varphi. \nonumber
\end{eqnarray}
At last, curvatures for the extra fields look like:
\begin{eqnarray}
{\cal R}^{\alpha(4)} &=& D \Omega^{\alpha(4)} + \frac{a_0{}^2}{80} 
e^\alpha{}_{\dot\alpha} \Omega^{\alpha(3)\dot\alpha} - \frac{m^2}{2}
E^\alpha{}_\beta B^{\alpha(3)\beta} - \frac{\tilde{m}a_0}{60}
E^{\alpha(2)} B^{\alpha(2)}, \nonumber \\
{\cal B}^{\alpha(4)} &=& D B^{\alpha(4)} - \Omega^{\alpha(4)}
+ \frac{a_0{}^2}{80} e^\alpha{}_{\dot\alpha} B^{\alpha(3)\dot\alpha}, 
\\
{\cal B}^{\alpha(3)\dot\alpha} &=& D B^{\alpha(3)\dot\alpha} 
- \Omega^{\alpha(3)\dot\alpha} + e_\beta{}^{\dot\alpha}
B^{\alpha(3)\beta} + \frac{\tilde{m}}{6} e^{\alpha\dot\alpha}
B^{\alpha(2)} + \frac{M^2}{6} e^\alpha{}_{\dot\beta} 
\pi^{\alpha(2)\dot\alpha\dot\beta}, \nonumber \\
\Pi^{\alpha(2)\dot\alpha(2)} &=& D \pi^{\alpha(2)\dot\alpha(2)}
- f^{\alpha(2)\dot\alpha(2)} + e_\beta{}^{\dot\alpha}
B^{\alpha(2)\beta\dot\alpha} + e^\alpha{}_{\dot\beta}
B^{\alpha\dot\alpha(2)\dot\beta} + \frac{\tilde{m}}{3}
e^{\alpha\dot\alpha} \pi^{\alpha\dot\alpha}. \nonumber
\end{eqnarray}

In the constructive approach, based on the metric-like formalism, the
two notions "on-shell" and "on free equations" are
equivalent. However, due to the presence of the extra fields in the
frame-like formalism such equivalence does not apply. In this case,
 "on-shell" implies that all gauge invariant curvatures of
physical and auxiliary fields vanish, leaving only four
curvatures ${\cal R}^{\alpha(4)}$, ${\cal B}^{\alpha(4)}$, 
${\cal B}^{\alpha(3)\dot\alpha}$ and $\Pi^{\alpha(2)\dot\alpha(2)}$:
\begin{eqnarray}
{\cal R}^{\alpha(4)} &\approx& E_{\beta(2)} W^{\alpha(4)\beta(2)},
\qquad {\cal B}^{\alpha(4)} \approx e_{\beta\dot\alpha}
B^{\alpha(4)\beta\dot\alpha}, \nonumber \\
{\cal B}^{\alpha(3)\dot\alpha} &\approx& e_{\beta\dot\beta}
B^{\alpha(3)\beta\dot\alpha\dot\beta},
\qquad \Pi^{\alpha(2)\dot\alpha(2)} \approx e_{\beta\dot\beta}
\pi^{\alpha(2)\beta\dot\alpha(2)\dot\beta}, 
\end{eqnarray}
where gauge invariant zero-forms $W^{\alpha(6)}$,
$B^{\alpha(5)\dot\alpha}$, $B^{\alpha(4)\dot\alpha(2)}$ and
$\pi^{\alpha(3)\dot\alpha(3)}$ are generalizations of the Weyl tensor
in gravity. In what follows it will be important that our four gauge
invariant curvatures satisfy the following differential identities:
\begin{eqnarray}
D {\cal R}^{\alpha(4)} &\approx& - \frac{m^2}{2} E^\alpha{}_\beta
{\cal B}^{\alpha(3)\beta}, \nonumber \\
D {\cal B}^{\alpha(4)} &\approx& - {\cal R}^{\alpha(4)} -
\frac{a_0{}^2}{80} e^\alpha{}_{\dot\alpha} 
{\cal B}^{\alpha(3)\dot\alpha}, \\
D {\cal B}^{\alpha(3)\dot\alpha} &\approx& - e_\beta{}^{\dot\alpha}
{\cal B}^{\alpha(3)\beta} - \frac{M^2}{6} e^\alpha{}_{\dot\beta}
\Pi^{\alpha(2)\dot\alpha\dot\beta}, \nonumber \\
D \Pi^{\alpha(2)\dot\alpha(2)} &\approx& - e_\beta{}^{\dot\alpha}
{\cal B}^{\alpha(2)\beta\dot\alpha} - e^\alpha{}_{\dot\beta}
{\cal B}^{\alpha\dot\alpha(2)\dot\beta}, \nonumber
\end{eqnarray}
as well as a number of algebraic ones:
\begin{eqnarray}
 0 &\approx& e_\beta{}^{\dot\alpha} 
{\cal R}^{\alpha(2)\beta\dot\alpha} + e^\alpha{}_{\dot\beta} 
{\cal R}^{\alpha\dot\alpha(2)\dot\beta}, \qquad 0 \approx 
e_\beta{}^{\dot\alpha} {\cal R}^{\alpha\beta} + e^\alpha{}_{\dot\beta}
{\cal R}^{\dot\alpha\dot\beta}, \nonumber \\
0 &\approx&   E_{\alpha(2)} {\cal B}^{\alpha(2)} + E_{\dot\alpha(2)}
{\cal B}^{\dot\alpha(2)}, \qquad 0 \approx e_{\alpha\dot\alpha}
\Pi^{\alpha\dot\alpha}. 
\end{eqnarray}

It follows from the relations (\ref{unitary}) that, in $dS_4$ space,
there exists a unitary forbidden region with $m^2 < 6\Lambda$ for
massive spin 3. At the boundary of this region at $m^2 = 6\Lambda$ we
discover a first partially massless case, while inside the region at
$m^2 = 4\Lambda$, we find a second partially massless case. Note,
that both cases are unitary.
\begin{figure}[htb]
\setlength{\unitlength}{0.8mm}
\begin{center}
\begin{picture}(80,60)
\put(10,20){\vector(1,0){60}}
\put(10,0){\vector(0,1){50}}
\put(0,45){\makebox(10,10)[]{$\Lambda$}}
\put(70,15){\makebox(10,10)[]{$m^2$}}

\put(10,20){\line(2,1){60}}
\put(10,20){\line(1,1){30}}
\put(38,50){\makebox(10,10)[]{$m^2=4\Lambda$}}
\put(68,50){\makebox(10,10)[]{$m^2=6\Lambda$}}
\end{picture}
\caption{Unitary forbidden region $m^2 < 6\Lambda$ for massive spin 3}
\end{center}
\end{figure}
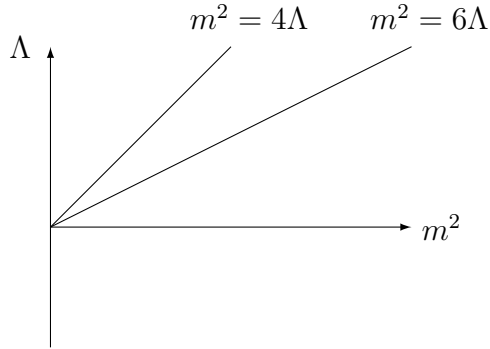

\subsection{First partially massless limit}

This limit corresponds to
\begin{equation}
m^2 = 6\Lambda \quad \Rightarrow \quad M^2 = 0, \qquad
a_0{}^2 = 40\Lambda, \qquad \tilde{a}_0 = 0.
\end{equation}
In this limit helicity 0 decouples leaving us with the Lagrangian
\begin{eqnarray}
{\cal L}_0 &=& - 3 \Omega^{\alpha(2)\beta\dot\alpha}
E_\beta{}^\gamma \Omega_{\alpha(2)\gamma\dot\alpha} + 
\Omega^{\alpha(3)\dot\alpha} E_{\dot\alpha}{}^{\dot\beta}
\Omega^{\alpha(3)\dot\beta} - 2 \Omega^{\alpha(2)\beta\dot\alpha}
e_\beta{}^{\dot\beta} D f_{\alpha(2)\dot\alpha\dot\beta} \nonumber \\
 && + 2 \Omega^{\alpha\beta} E_\beta{}^\gamma \Omega_{\alpha\gamma}
+ 2 \Omega^{\alpha\beta} e_\beta{}^{\dot\alpha} D f_{\alpha\dot\alpha}
 + 4E B_{\alpha(2)} B^{\alpha(2)} + 2 E_{\alpha(2)} B^{\alpha(2)}
D A \nonumber \\
 && + 4\tilde{m} [\Omega^{\alpha\beta(2)\dot\alpha} E_{\beta(2)}
f_{\alpha\dot\alpha} + \frac{1}{3} f^{\alpha(2)\dot\alpha(2)}
E_{\beta(2)} \Omega_{\dot\alpha(2)} ] \nonumber \\
 && + a_0 [ \Omega^{\alpha(2)} E_{\alpha(2)} A - 2 B^{\alpha\beta}
E_\beta{}^{\dot\alpha} f_{\alpha\dot\alpha}] + h.c.   
\end{eqnarray}
Moreover, the three fields $\pi^{\alpha(2)\dot\alpha(2)}$, 
$\pi^{\alpha\dot\alpha}$ and $\varphi$ completely decouple from all
the gauge invariant curvatures. In particular, this means that the
three gauge one-forms $f^{\alpha(2)\dot\alpha(2)}$,
$f^{\alpha\dot\alpha}$ and $A$ lose their Stueckelberg zero-forms and
some of the gauge symmetries remain unbroken (hence the name partially
massless field). 

In what follows, we will not consider this case separately. Instead,
we will discuss what happens in this limit when dealing with a
general massive case.

\subsection{Second partially massless limit}

This limit corresponds to
\begin{equation}
m^2 = 4\Lambda \quad \Rightarrow \quad M^2 = - 2\Lambda, \qquad
a_0 = 0, \qquad \tilde{m}^2 = 3\Lambda.
\end{equation}
In this limit helicities $(\pm 1, 0)$ decouple and the Lagrangian for
the remaining helicities $(\pm 3, \pm 2)$ has the form
\begin{eqnarray}
{\cal L}_0 &=& - 3 \Omega^{\alpha(2)\beta\dot\alpha}
E_\beta{}^\gamma \Omega_{\alpha(2)\gamma\dot\alpha} + 
\Omega^{\alpha(3)\dot\alpha} E_{\dot\alpha}{}^{\dot\beta}
\Omega^{\alpha(3)\dot\beta} - 2 \Omega^{\alpha(2)\beta\dot\alpha}
e_\beta{}^{\dot\beta} D f_{\alpha(2)\dot\alpha\dot\beta} \nonumber \\
 && + 2 \Omega^{\alpha\beta} E_\beta{}^\gamma \Omega_{\alpha\gamma}
+ 2 \Omega^{\alpha\beta} e_\beta{}^{\dot\alpha} f_{\alpha\dot\alpha} 
\nonumber \\
 && + 4\tilde{m} [\Omega^{\alpha\beta(2)\dot\alpha} E_{\beta(2)}
f_{\alpha\dot\alpha} + \frac{1}{3} f^{\alpha(2)\dot\alpha(2)}
E_{\beta(2)} \Omega_{\dot\alpha(2)}] \nonumber \\
 && - M^2 f^{\alpha\beta\dot\alpha(2)} E_\beta{}^\gamma
f_{\alpha\gamma\dot\alpha(2)} + 4M^2 f^{\alpha\dot\alpha}
E_\alpha{}^\beta f_{\beta\dot\alpha} + h.c.  
\end{eqnarray}
This Lagrangian is invariant under the following gauge
transformations:
\begin{eqnarray}
\delta \Omega^{\alpha(3)\dot\alpha} &=& D \eta^{\alpha(3)\dot\alpha}
+ e_\beta{}^{\dot\alpha} \eta^{\alpha(3)\beta} 
 + \frac{\tilde{m}}{6} e^{\alpha\dot\alpha} \eta^{\alpha(2)} +
\frac{M^2}{6} e^\alpha{}_{\dot\beta}
\xi^{\alpha(2)\dot\alpha\dot\beta}, \nonumber \\
\delta f^{\alpha(2)\dot\alpha(2)} &=& D \xi^{\alpha(2)\dot\alpha(2)}
 + e_\beta{}^{\dot\alpha} \eta^{\alpha(2)\beta\dot\alpha}
+  e^\alpha{}_{\dot\beta} \eta^{\alpha\dot\alpha(2)\dot\beta} 
+ \frac{\tilde{m}}{3} e^{\alpha\dot\alpha} \xi^{\alpha\dot\alpha},
\nonumber \\
\delta \Omega^{\alpha(2)} &=& D \eta^{\alpha(2)} + 2\tilde{m}
e_{\beta\dot\alpha} \eta^{\alpha(2)\beta\dot\alpha} + M^2
e^\alpha{}_{\dot\alpha} \xi^{\alpha\dot\alpha}, \\
\delta f^{\alpha\dot\alpha} &=& D \xi^{\alpha\dot\alpha} +
e_\beta{}^{\dot\alpha} \eta^{\alpha\beta} + e^\alpha{}_{\dot\beta}
\eta^{\dot\alpha\dot\beta} + \frac{2\tilde{m}}{3} e_{\beta\dot\beta}
\xi^{\alpha\beta\dot\alpha\dot\beta}. \nonumber
\end{eqnarray}
The gauge invariant curvatures for the Lagrangian fields look like
\begin{eqnarray}
{\cal R}^{\alpha(3)\dot\alpha} &=& D \Omega^{\alpha(3)\dot\alpha}
 + e_\beta{}^{\dot\alpha} \Omega^{\alpha(3)\beta} 
+ \frac{\tilde{m}}{6}e^{\alpha\dot\alpha} \Omega^{\alpha(2)} +
\frac{M^2}{6} e^\alpha{}_{\dot\beta} f^{\alpha(2)\dot\alpha\dot\beta},
\nonumber \\
{\cal T}^{\alpha(2)\dot\alpha(2)} &=& D f^{\alpha(2)\dot\alpha(2)}
+ e_\beta{}^{\dot\alpha} \Omega^{\alpha(2)\beta\dot\alpha}
+ e^\alpha{}_{\dot\beta} \Omega^{\alpha\dot\alpha(2)\dot\beta}
+ \frac{\tilde{m}}{3} e^{\alpha\dot\alpha} f^{\alpha\dot\alpha}, \\
{\cal R}^{\alpha(2)} &=& D \Omega^{\alpha(2)} + 2\tilde{m}
e_{\beta\dot\beta} \Omega^{\alpha(2)\beta\dot\beta} + M^2
e^\alpha{}_{\dot\alpha} f^{\alpha\dot\alpha} - 4\tilde{m} E_{\beta(2)}
B^{\alpha(2)\beta(2)}, \nonumber \\
{\cal T}^{\alpha\dot\alpha} &=& D f^{\alpha\dot\alpha} 
+ e_\beta{}^{\dot\alpha} \Omega^{\alpha\beta} + e^\alpha{}_{\dot\beta}
\Omega^{\dot\alpha\dot\beta} + \frac{2\tilde{m}}{3} e_{\beta\dot\beta}
f^{\alpha\beta\dot\alpha\dot\beta}, \nonumber
\end{eqnarray}
while for the extra fields we have
\begin{equation}
{\cal R}^{\alpha(4)} = D \Omega^{\alpha(4)} + 2\lambda^2 
E^\alpha{}_\beta B^{\alpha(3)\beta}, \qquad {\cal B}^{\alpha(4)} = D
B^{\alpha(4)} - \Omega^{\alpha(4)}.
\end{equation}
In \cite{SV06} Skvortsov and Vasiliev proposed a very simple and
convenient formalism for describing bosonic partially massless
fields. In our framework, this corresponds to a partial gauge fixing
where one sets $B^{\alpha(4)} = 0$ and solves its equation. After
that, the Lagrangian can be expressed in a simple and explicitly gauge
invariant form:
\begin{equation}
{\cal L}_0 = \frac{1}{\Lambda} {\cal R}_{\alpha(3)\dot\alpha} 
{\cal R}^{\alpha(3)\dot\alpha} - \frac{1}{4\Lambda} 
{\cal R}_{\alpha(2)} {\cal R}^{\alpha(2)} + h.c.
\end{equation}

\subsection{Massless limit}

In this case the Lagrangian has the form
\begin{eqnarray}
 {\cal L}_0 &=& - 3 \Omega^{\alpha(2)\beta\dot\alpha}
E_\beta{}^\gamma \Omega_{\alpha(2)\gamma\dot\alpha} + 
\Omega^{\alpha(3)\dot\alpha} E_{\dot\alpha}{}^{\dot\beta}
\Omega^{\alpha(3)\dot\beta} - 2 \Omega^{\alpha(2)\beta\dot\alpha}
e_\beta{}^{\dot\beta} D f_{\alpha(2)\dot\alpha\dot\beta} \nonumber \\
 && - 6\lambda^2 f^{\alpha\beta\dot\alpha(2)} E_\beta{}^\gamma
f_{\alpha\gamma\dot\alpha(2)}  + h.c. 
\end{eqnarray}
and is invariant under the following gauge transformations
\begin{eqnarray}
\delta \Omega^{\alpha(3)\dot\alpha} &=& D \eta^{\alpha(3)\dot\alpha}
+ e_\beta{}^{\dot\alpha} \eta^{\alpha(3)\beta} + \lambda^2
e^\alpha{}_{\dot\beta} \xi^{\alpha(2)\dot\alpha\dot\beta}, \nonumber 
\\
\delta f^{\alpha(2)\dot\alpha(2)} &=& D \xi^{\alpha(2)\dot\alpha(2)}
+ e_\beta{}^{\dot\alpha} \eta^{\alpha(2)\beta\dot\alpha}
+ e^\alpha{}_{\dot\beta} \eta^{\alpha\dot\alpha(2)\dot\beta}.
\end{eqnarray}
A complete set of the gauge invariant curvatures contains
\begin{eqnarray}
{\cal R}^{\alpha(4)} &=& D \Omega^{\alpha(4)} + \lambda^2 
e^\alpha{}_{\dot\alpha} \Omega^{\alpha(3)\dot\alpha}, \nonumber \\
{\cal R}^{\alpha(3)\dot\alpha} &=& D \Omega^{\alpha(3)\dot\alpha}
 + e_\beta{}^{\dot\alpha} \Omega^{\alpha(3)\beta} 
+ \lambda^2 e^\alpha{}_{\dot\beta} f^{\alpha(2)\dot\alpha\dot\beta}, 
\\
{\cal T}^{\alpha(2)\dot\alpha(2)} &=& D f^{\alpha(2)\dot\alpha(2)}
+ e_\beta{}^{\dot\alpha} \Omega^{\alpha(2)\beta\dot\alpha}
+ e^\alpha{}_{\dot\beta} \Omega^{\alpha\dot\alpha(2)\dot\beta},
\nonumber
\end{eqnarray}
while the Lagrangian can be written as follows:
\begin{equation}
{\cal L}_0 = - \frac{1}{12\lambda^4} {\cal R}_{\alpha(4)} 
{\cal R}^{\alpha(4)} - \frac{1}{3\lambda^2} 
{\cal R}_{\alpha(3)\dot\alpha} {\cal R}^{\alpha(3)\dot\alpha} + h.c.
\end{equation}

\section{Massless case}

In four dimensions, there are two types of cubic vertices for
massless integer spins \cite{Met18a,Met22}. One type belongs to
the so-called trivially invariant vertices, with a number of
derivatives $N_B = s_1 + s_2 + s_3$. In \cite{Zin24a}, it was shown
that if the so-called triangular inequality  $s_1 \le s_2 + s_3$
(here we assume that $s_1 \ge s_2 \ge s_3$) holds, such vertices can
be expressed using gauge invariant zero-forms. For the case
$(3,3,2)$, the corresponding vertex can be written as two on-shell
equivalent forms:
\begin{equation}
{\cal L} \sim E W^{\alpha(2)\beta(2)} W_{\alpha(2)\gamma(4)}
W_{\beta(2)}{}^{\gamma(4)} \approx W^{\alpha(2)\beta(2)}
{\cal R}_{\alpha(2)\gamma(2)} {\cal R}_{\beta(2)}{}^{\gamma(2)},
\end{equation}
where
$$
R^{\alpha(2)} \approx E_{\beta(2)} W^{\alpha(2)\beta(2)}.
$$
For the vertices of the second type the number of derivatives is 
$N_B = s_1 + s_2 - s_3$. Moreover, if the so-called strict
triangular inequality $s_1 < s_2 + s_3$ holds, the corresponding
vertex appears to be non-abelian.

To construct a non-abelian vertex $(3,3,2)$ we use the 
Fradkin-Vasiliev formalism 
\cite{FV87,FV87a,Vas11,KhZ20a,Zin24a}\footnote{Previous work in direct
constructive approach see \cite{Zin08,Zin10}}.
Recall that the first step is to find a consistent deformation of all
gauge invariant curvatures. Here, "consistency" means that the
deformed curvatures $\hat{\cal R} = {\cal R} + \Delta {\cal R}$
transform covariantly $\delta \hat{\cal R} \sim {\cal R}$. For the
spin 3 we obtain (we set the coupling constant to be 1):
\begin{eqnarray}
\Delta {\cal R}^{\alpha(4)} &=&  \omega^\alpha{}_\beta
\Omega^{\alpha(3)\beta} + \lambda^2 h^\alpha{}_{\dot\alpha}
\Omega^{\alpha(3)\dot\alpha}, \nonumber \\
\Delta {\cal R}^{\alpha(3)\dot\alpha} &=& \omega^\alpha{}_\beta
\Omega^{\alpha(2)\beta\dot\alpha} +  
\omega^{\dot\alpha}{}_{\dot\beta} \Omega^{\alpha(3)\dot\beta}
+ h_\beta{}^{\dot\alpha} \Omega^{\alpha(3)\beta} + \lambda^2
h^\alpha{}_{\dot\beta} f^{\alpha(2)\dot\alpha\dot\beta}, \\
\Delta {\cal T}^{\alpha(2)\dot\alpha} &=& 
\omega^\alpha{}_\beta f^{\alpha\beta\dot\alpha(2)} + 
\omega^{\dot\alpha}{}_{\dot\beta} f^{\alpha(2)\dot\alpha\dot\beta}
+ h_\beta{}^{\dot\alpha} \Omega^{\alpha(2)\beta\dot\alpha}
+  h^\alpha{}_{\dot\beta} \Omega^{\alpha\dot\alpha(2)\dot\beta},
\nonumber
\end{eqnarray}
which corresponds to the standard substitution rules. For the graviton
we obtain:
\begin{eqnarray}
\delta R^{\alpha(2)} &=& b_1 [\Omega^{\alpha\beta(3)} 
\Omega^\alpha{}_{\beta(3)} + 3\lambda^2 
\Omega^{\alpha\beta(2)\dot\alpha} \Omega^\alpha{}_{\beta(2)\dot\alpha}
+ 6\lambda^4 f^{\alpha\beta\dot\alpha(2)} 
f^\alpha{}_{\beta\dot\alpha(2)} + \lambda^2
\Omega^{\alpha\dot\alpha(3)} \Omega^\alpha{}_{\dot\alpha(3)}],
\nonumber \\ 
\Delta T^{\alpha\dot\alpha} &=& b_1 [ 2 \Omega^{\alpha\beta(3)}
\Omega_{\beta(3)}{}^{\dot\alpha} +  6\lambda^2 
\Omega^{\alpha\beta(2)\dot\beta} f_{\beta(2)\dot\beta}{}^{\dot\alpha}
+ 6\lambda^2 f^{\alpha\beta\dot\beta(2)} 
\Omega_{\beta\dot\beta(2)}{}^{\dot\alpha} + 2 
\Omega^{\alpha\dot\beta(3)} \Omega_{\dot\beta(3)}{}^{\dot\alpha} ].
\end{eqnarray}
Now we consider a deformed Lagrangian, i.e. the sum of the free
Lagrangians with the deformed curvatures instead of the initial ones:
\begin{equation}
\hat{\cal L} = - \frac{1}{12\lambda^4} \hat{\cal R}_{\alpha(4)}
\hat{\cal R}^{\alpha(4)} - \frac{1}{3\lambda^2}
\hat{\cal R}_{\alpha(3)\dot\alpha} \hat{\cal R}^{\alpha(3)\dot\alpha}
+ \frac{1}{4\lambda^2} \hat{R}_{\alpha(2)} \hat{R}^{\alpha(2)} + h.c.
\end{equation}
Non vanishing on-shell variations
\begin{equation}
\delta \hat{\cal R}^{\alpha(4)} \approx R^\alpha{}_\beta
\eta^{\alpha(3)\beta}, \qquad \delta \hat{R}^{\alpha(2)} \approx
2b_1 {\cal R}^{\alpha\beta(3)} \eta^\alpha{}_{\beta(3)}
\end{equation}
produce
\begin{equation}
\delta \hat{\cal L} \approx \frac{2}{\lambda^2} 
[b_1 - \frac{1}{3\lambda^2}] {\cal R}_{\alpha(3)\beta} 
R^\beta{}_\gamma \eta^{\alpha(3)\gamma},
\end{equation}
so we put
\begin{equation}
b_1 = \frac{1}{3\lambda^2}.
\end{equation}
At last, we extract a cubic part of the deformed Lagrangian
\begin{eqnarray}
{\cal L}_1 &=& - \frac{1}{6\lambda^4} {\cal R}_{\alpha(4)}
[  \omega^\alpha{}_\beta \Omega^{\alpha(3)\beta} + \lambda^2
h^\alpha{}_{\dot\alpha} \Omega^{\alpha(3)\dot\alpha} ] \nonumber \\
 && - \frac{2}{3\lambda^2} {\cal R}_{\alpha(3)\dot\alpha}
 [ \omega^\alpha{}_\beta \Omega^{\alpha(2)\beta\dot\alpha} +  
\omega^{\dot\alpha}{}_{\dot\beta} \Omega^{\alpha(3)\dot\beta}
+ h_\beta{}^{\dot\alpha} \Omega^{\alpha(3)\beta} + \lambda^2
h^\alpha{}_{\dot\beta} f^{\alpha(2)\dot\alpha\dot\beta} ] \nonumber \\
 && + \frac{1}{6\lambda^4} R_{\alpha(2)} [ \Omega^{\alpha\beta(3)} 
\Omega^\alpha{}_{\beta(3)} + 3\lambda^2 
\Omega^{\alpha\beta(2)\dot\alpha} \Omega^\alpha{}_{\beta(2)\dot\alpha}
\nonumber \\
 && \qquad \qquad + 3\lambda^4 f^{\alpha\beta\dot\alpha(2)} 
f^\alpha{}_{\beta\dot\alpha(2)} + \lambda^2
\Omega^{\alpha\dot\alpha(3)} \Omega^\alpha{}_{\dot\alpha(3)} ] + h.c. 
\end{eqnarray}
Formally, this expression contains terms with up to six derivatives,
but these terms form a total derivative and can be ignored. Thus, the
vertex has up to four derivatives as it should \footnote{For the
three arbitrary spins $s_{1,2,3}$ the Fradkin-Vasiliev formalism
generates terms with up to $s_1+s_2+s_3-2$ derivatives, but it was
shown in \cite{KhZ20a} that all terms with more derivatives than
$s_1+s_2-s_3$ form total derivatives or vanish on-shell.}. Using
explicit expressions for the curvatures and integrating by parts, one
can show that the vertex can be rewritten in an on-shell equivalent
form:
\begin{eqnarray}
{\cal L}_1 &=& \frac{2}{3\lambda^2} R_{\alpha\beta}
\Omega^{\alpha\dot\alpha(3)} \Omega^\beta{}_{\dot\alpha(3)}
 + 2 R_{\alpha\beta} f^{\alpha\gamma\dot\alpha(2)}
f^\beta{}_{\gamma\dot\alpha(2)} \nonumber \\
 && - 2 D \Omega_{\alpha(2)\beta\dot\alpha} h^\beta{}_{\dot\beta}
f^{\alpha(2)\dot\alpha\dot\beta} 
+ 2 [ \omega_\alpha{}^\beta \Omega_{\alpha\gamma\beta\dot\alpha}
+ \omega_\gamma{}^\beta \Omega_{\alpha(2)\beta\dot\alpha}
+ \omega_{\dot\alpha}{}^{\dot\beta} \Omega_{\alpha(2)\gamma\dot\beta}]
e^\gamma{}_{\dot\gamma} f^{\alpha(2)\dot\alpha\dot\gamma} \nonumber \\
 && - 3 \Omega_{\alpha(2)\beta\dot\alpha} e^\beta{}_{\dot\beta}
h^{\gamma\dot\beta} \Omega_\gamma{}^{\alpha(2)\dot\alpha}
+ \Omega_{\alpha(3)\dot\alpha} e_\beta{}^{\dot\alpha}
h^{\beta\dot\beta} \Omega^{\alpha(3)}{}_{\dot\beta} 
 - 6\lambda^2 f_{\alpha\beta\dot\alpha(2)} e^\beta{}_{\dot\beta}
h^{\gamma\dot\beta} f_\gamma{}^{\alpha\dot\alpha(2)} + h.c.
\end{eqnarray}
The first line contains non-minimal corrections, while two other lines
exactly correspond to standard covariantization of the free
Lagrangian:
\begin{equation}
e^{\alpha\dot\alpha} \Rightarrow e^{\alpha\dot\alpha} +
h^{\alpha\dot\alpha}, \qquad D \Rightarrow D + \omega^{\alpha(2)}
L_{\alpha(2)} + \omega^{\dot\alpha(2)} L_{\dot\alpha(2)},
\end{equation}
where $L_{\alpha(2)}$, $L_{\dot\alpha(2)}$ are Lorentz group
generators.

Recall that we set the gravitational coupling constant equal to 1, and
as a result, the  coefficient of non-minimal interactions appears to
be singular in the flat limit. But we can perform a rescaling so that
the flat limit becomes possible, leaving us with only the non-minimal
terms. A similar rescaling is also possible for three arbitrary
massless fields \cite{KhZ20a}. Moreover, there exists a one-to-one
correspondence between these limits and the cubic vertices constructed
directly in flat space \cite{Met18a}.

\section{Partially massless case}

The simplest and most straightforward way to construct a minimal
vertex (containing standard minimal interactions and non-minimal ones
with the minimum number of derivatives) is to use the
Skvortsov-Vasiliev description of partially massless spin 3
\cite{SV06}. This description only contains one-forms, so the
Fradkin-Vasiliev formalism works exactly as in the massless case
without any ambiguities related with field redefinitions.

We begin with the consistent deformations for all gauge  invariant
curvatures. For the partially massless spin 3 we obtain:
\begin{eqnarray}
\Delta {\cal R}^{\alpha(3)\dot\alpha} &=& \omega^\alpha{}_\beta
\Omega^{\alpha(2)\beta\dot\alpha} + \omega^{\dot\alpha}{}_{\dot\beta}
\Omega^{\alpha(3)\dot\beta} + \frac{\tilde{m}}{6} h^{\alpha\dot\alpha}
\Omega^{\alpha(2)} + \frac{\lambda^2}{3} h^\alpha{}_{\dot\beta}
f^{\alpha(2)\dot\alpha\dot\beta}, \nonumber \\
\Delta {\cal R}^{\alpha(2)} &=& \omega^\alpha{}_\beta
\Omega^{\alpha\beta} + 2\tilde{m} h_{\beta\dot\beta} 
\Omega^{\alpha(2)\beta\dot\beta} + 2\lambda^2 h^\alpha{}_{\dot\alpha}
f^{\alpha\dot\alpha}, 
\end{eqnarray}
which again correspond to the standard minimal substitution rules.
And for the graviton we get
\begin{eqnarray}
\Delta R^{\alpha(2)} &=& b_1 [\Omega^{\alpha\beta(2)\dot\alpha}
\Omega^\alpha{}_{\beta(2)\dot\alpha} + \frac{\lambda^2}{3}
f^{\alpha\beta\dot\alpha(2)} f^\alpha{}_{\beta\dot\alpha(2)}
+ \frac{1}{3} \Omega^{\alpha\dot\alpha(3)} 
\Omega^\alpha{}_{\dot\alpha(3)} \nonumber \\
 && - \frac{1}{6} \Omega^{\alpha\beta} \Omega^\alpha{}_\beta 
- \frac{\lambda^2}{3} f^{\alpha\dot\alpha} f^\alpha{}_{\dot\alpha}], 
\\
\Delta T^{\alpha\dot\alpha} &=& b_1 [ \frac{2}{3} 
\Omega^{\alpha\beta(2)\dot\beta}f_{\beta(2)\dot\beta}{}^{\dot\alpha} 
- \frac{1}{\tilde{m}} \Omega^{\alpha\beta(2)\dot\alpha}
\Omega_{\beta(2)} - \frac{2}{3} 
\Omega^{\alpha\beta} f_\beta{}^{\dot\alpha} + h.c. + \tilde{m}
f^{\alpha\beta\dot\alpha\dot\beta} f_{\beta\dot\beta} ]. \nonumber
\end{eqnarray}
Now we consider a deformed Lagrangian
\begin{equation}
\hat{\cal L} = - \frac{1}{\lambda^2} 
\hat{\cal R}_{\alpha(3)\dot\alpha} \hat{\cal R}^{\alpha(3)\dot\alpha}
+ \frac{1}{4\lambda^2} \hat{\cal R}_{\alpha(2)}
\hat{\cal R}^{\alpha(2)} + \frac{1}{4\lambda^2} \hat{R}_{\alpha(2)}
\hat{R}^{\alpha(2)} + h.c. 
\end{equation}
and require it to be gauge invariant. Non vanishing on-shell
variations
\begin{eqnarray}
\delta \hat{\cal R}^{\alpha(3)\dot\alpha} &=& R^\alpha{}_\beta
\eta^{\alpha(2)\beta\dot\alpha}, \qquad
\delta \hat{\cal R}^{\alpha(2)} = R^\alpha{}_\beta \eta^{\alpha\beta},
\nonumber \\
\delta \hat{R}^{\alpha(2)} &=& 2b_1 
{\cal R}^{\alpha\beta(2)\dot\alpha} \eta^\alpha{}_{\beta(2)\dot\alpha}
- \frac{b_1}{3} {\cal R}^{\alpha\beta} \eta^\alpha{}_\beta
\end{eqnarray}
produce
\begin{equation}
\delta{\cal L} \approx \frac{2(b_1-3)}{\lambda^2} 
[ {\cal R}_{\alpha(2)\beta\dot\alpha} R^\beta{}_\gamma
\eta^{\alpha(2)\gamma\dot\alpha} - \frac{1}{6} {\cal R}_{\alpha\beta}
R^\beta{}_\gamma \eta^{\alpha\gamma}], 
\end{equation}
so we must put
\begin{equation}
b_1 = 3.
\end{equation}

At last we extract the cubic part of the deformed Lagrangian:
\begin{eqnarray}
{\cal L}_1 &=& - \frac{2}{\lambda^2} {\cal R}_{\alpha(3)\dot\alpha}
[ \omega^\alpha{}_\beta \Omega^{\alpha(2)\beta\dot\alpha} +
\omega^{\dot\alpha}{}_{\dot\beta} \Omega^{\alpha(3)\dot\beta} +
\frac{\tilde{m}}{6} h^{\alpha\dot\alpha} \Omega^{\alpha(2)} +
\frac{\lambda^2}{3} h^\alpha{}_{\dot\beta}
f^{\alpha(2)\dot\alpha\dot\beta}  ] \nonumber \\
 && + \frac{1}{2\lambda^2} {\cal R}_{\alpha(2)} 
[ \omega^\alpha{}_\beta \Omega^{\alpha\beta} + 2\tilde{m}
h_{\beta\dot\beta} \Omega^{\alpha(2)\beta\dot\beta} + 2\lambda^2
h^\alpha{}_{\dot\alpha} f^{\alpha\dot\alpha}   ] \nonumber \\
 && + \frac{1}{2\lambda^2} R_{\alpha(2)} 
[3 \Omega^{\alpha\beta(2)\dot\alpha}
\Omega^\alpha{}_{\beta(2)\dot\alpha} + \lambda^2
f^{\alpha\beta\dot\alpha(2)} f^\alpha{}_{\beta\dot\alpha(2)}
+ \Omega^{\alpha\dot\alpha(3)} \Omega^\alpha{}_{\dot\alpha(3)}
\nonumber \\
 && \qquad \qquad - \frac{1}{2} \Omega^{\alpha\beta} 
\Omega^\alpha{}_\beta - \lambda^2 f^{\alpha\dot\alpha} 
f^\alpha{}_{\dot\alpha}] + h.c.
\end{eqnarray}
Contrary to the massless case there are no terms with six derivatives
here. By the same technique as before we have managed to transform
this vertex into the following suggestive form:
\begin{eqnarray}
{\cal L}_1 &=& \frac{1}{2\lambda^2} R_{\alpha(2)}
[ 2 \Omega^{\alpha\dot\alpha(3)} \Omega^\alpha{}_{\dot\alpha(3)}
+ \lambda^2 f^{\alpha\beta\dot\alpha(2)} 
f^\alpha{}_{\beta\dot\alpha(2)} - \lambda^2 f^{\alpha\dot\alpha}
f^\alpha{}_{\dot\alpha}] \nonumber \\
 && - 2 D \Omega_{\alpha(2)\beta\dot\alpha} h^\beta{}_{\dot\beta}
f^{\alpha(2)\dot\alpha\dot\beta} + 2 [\omega^\alpha{}_\gamma
\Omega^{\alpha\beta\gamma\dot\alpha}
+ \omega^\beta{}_\gamma \Omega^{\alpha(2)\gamma\dot\alpha}
 + \omega^{\dot\alpha}{}_{\dot\gamma} 
\Omega^{\alpha(2)\beta\dot\gamma} ] e_\beta{}^{\dot\beta} 
f_{\alpha(2)\dot\alpha\dot\beta} \nonumber \\ 
 && + 2 D \Omega_{\alpha\beta} h^\beta{}_{\dot\alpha}
f^{\alpha\dot\alpha}  - 2 [ \omega^\alpha{}_\gamma
\Omega^{\beta\gamma} + \omega^\beta{}_\gamma \Omega^{\alpha\gamma}]
e_\beta{}^{\dot\alpha} f_{\alpha\dot\alpha} \nonumber \\ 
 && - 3 \Omega_{\alpha(2)\beta\dot\alpha} e^\beta{}_{\dot\beta}
h^{\gamma\dot\beta} \Omega_\gamma{}^{\alpha(2)\dot\alpha} 
 + \Omega_{\alpha(3)\dot\alpha} e_\beta{}^{\dot\alpha}
h^{\beta\dot\beta} \Omega^{\alpha(3)\dot\beta} 
 + 2 \Omega_{\alpha\beta} e^\beta{}_{\dot\alpha}
h^{\gamma\dot\alpha} \Omega^\alpha{}_\gamma \nonumber \\
 && + 2\tilde{m} \Omega_{\alpha\beta(2)\dot\alpha} 
e^\beta{}_{\dot\beta} h^{\beta\dot\beta} f^{\alpha\dot\alpha} -
\frac{2\tilde{m}}{3} f_{\alpha(2)\dot\alpha(2)} e_\beta{}^{\dot\alpha}
h^{\beta\dot\alpha} \Omega^{\alpha(2)} \nonumber \\
 && - 2\lambda^2 f_{\alpha\beta\dot\alpha(2)} e^\beta{}_{\dot\beta}
h^{\gamma\dot\beta} f_\gamma{}^{\alpha\dot\alpha(2)} 
 + 8\lambda^2 f_{\alpha\dot\alpha} e^\beta{}_{\dot\beta}
h^{\gamma\dot\beta} f_\gamma{}^{\dot\alpha} + h.c., 
\end{eqnarray}
where the first line contains non-minimal terms, while all other lines
exactly correspond to the standard covariantization of the free
Lagrangian. 

\section{Massive case}

To construct a minimal gravitational vertex, we follow the same
down-up approach as in the massive spin 5/2 case in \cite{Zin25a}. 
Namely, we begin with the Lagrangian (\ref{lag_1}), (\ref{lag_2}),
(\ref{lag_3}), the gauge transformations (\ref{gaug_1}),
(\ref{gaug_2}) and gauge invariant curvatures (\ref{cur_1}),
(\ref{cur_2}), where the frame $e^{\alpha\dot\alpha}$ and Lorentz
covariant derivative are now dynamical. We still assume that torsion
is zero, so that the only source of Lagrangian non-invariance
is the non-commutativity of covariant derivatives. Then, by
straightforward calculations, we obtain the variations of the original
Lagrangian:
\begin{eqnarray}
\delta {\cal L}_0 &=& -2 [ R_\alpha{}^\gamma
\Omega_{\alpha\beta\gamma\dot\alpha} + R_\beta{}^\gamma
\Omega_{\alpha(2)\gamma\dot\alpha} + R_{\dot\alpha}{}^{\dot\gamma}
\Omega_{\alpha(2)\beta\dot\gamma}] e^\beta{}_{\dot\beta}
\xi^{\alpha(2)\dot\alpha\dot\beta} \nonumber \\
 && + 2 [ R_\alpha{}^\gamma f_{\alpha\gamma\dot\alpha\dot\beta}
+ R_{\dot\alpha}{}^{\dot\gamma} f_{\alpha(2)\dot\beta\dot\gamma}
+ R_{\dot\beta}{}^{\dot\gamma} f_{\alpha(2)\dot\alpha\dot\gamma}]
 e_\beta{}^{\dot\beta} \eta^{\alpha(2)\beta\dot\alpha} \nonumber \\
 && + 2 [ R_\alpha{}^\gamma \Omega_{\beta\gamma} + R_\beta{}^\gamma
\Omega_{\alpha\gamma}] e^\beta{}_{\dot\alpha} \xi^{\alpha\dot\alpha} 
\nonumber \\
 && - 2 [ R_\alpha{}^\gamma f_{\gamma\dot\alpha} +
R_{\dot\alpha}{}^{\dot\gamma} f_{\alpha\dot\gamma}]
e_\beta{}^{\dot\alpha} \eta^{\alpha\beta} + h.c.
\end{eqnarray}
Based on our experience on the partially massless case, we  try the
following ansatz for non-minimal terms:
\begin{equation}
{\cal L}_1 = R_{\alpha\beta} [\kappa_1 \Omega^{\alpha\dot\alpha(3)}
\Omega^\beta{}_{\dot\alpha(3)} + \kappa_2
f^{\alpha\gamma\dot\alpha(2)} f^\beta{}_{\gamma\dot\alpha(2)} +
\kappa_3 f^{\alpha\dot\alpha} f^\beta{}_{\dot\alpha} ] + h.c.
\end{equation}
Then, calculating variations of ${\cal L}_1$ under all gauge
transformations and using on-shell conditions, we found that if we
put
\begin{equation}
\kappa_1 = - \frac{4}{M^2}, \qquad \kappa_2 = - 1, \qquad
\kappa_3 = 1,
\end{equation}
then all variations $\delta ({\cal L}_0+ {\cal L}_1)$ can be
compensated by the following corrections for the graviton:
\begin{eqnarray}
\delta h^{\alpha\dot\alpha} &=& - \kappa_1 (\eta^{\alpha\beta(3)}
\Omega_{\beta(3)}{}^{\dot\alpha}  - \Omega^{\alpha\beta(3)}
\eta_{\beta(3)}{}^{\dot\alpha}) + 2 (\eta^{\alpha\beta(2)\dot\beta} 
 f_{\beta(2)\dot\beta}{}^{\dot\alpha} - 
\Omega^{\alpha\beta(2)\dot\beta} 
\xi_{\beta(2)\dot\beta}{}^{\dot\alpha}) \nonumber \\
 && - \frac{\tilde{m}\kappa_1}{2} (\eta^{\alpha\beta(2)\dot\alpha}
\Omega_{\beta(2)}  - \Omega^{\alpha\beta(2)\dot\alpha}
\eta_{\beta(2)}) - \frac{2\tilde{m}}{3}
(f^{\alpha\beta\dot\alpha\dot\beta} \xi_{\beta\dot\beta} -
\xi^{\alpha\beta\dot\alpha\dot\beta}  f_{\beta\dot\beta}) \nonumber \\
 && - 2 (\eta^{\alpha\beta} f_\beta{}^{\dot\alpha} -  
\Omega^{\alpha\beta} \xi_\beta{}^{\dot\alpha}) + \frac{a_0}{4}
(\xi^{\alpha\dot\alpha} A  - f^{\alpha\dot\alpha} \xi) + h.c.
\end{eqnarray}
Note that in the framework of Fradkin-Vasiliev formalism they
correspond to the following deformations of the torsion
\begin{eqnarray}
\Delta T^{\alpha\dot\alpha} &=& \kappa_1 \Omega^{\alpha\beta(3)}
\Omega_{\beta(3)}{}^{\dot\alpha} - 2 \Omega^{\alpha\beta(2)\dot\beta}
f_{\beta(2)\dot\beta}{}^{\dot\alpha} + \frac{\tilde{m}\kappa_1}{2}
\Omega^{\alpha\beta(2)\dot\alpha} \Omega_{\beta(2)} \nonumber \\
 && - \frac{2\tilde{m}}{3} f^{\alpha\beta\dot\alpha\dot\beta}
f_{\beta\dot\beta} + 2 \Omega^{\alpha\beta} f_\beta{}^{\dot\alpha}
- \frac{a_0}{4} f^{\alpha\dot\alpha} A + \dots + h.c.
\end{eqnarray}
where dots stand for the contributions of the zero-forms.

For the main coefficient at the non-minimal interactions we get
\begin{equation}
\kappa_1 = - \frac{4g}{\sqrt{m^2 - 6\Lambda}},
\end{equation}
where we restore the gravitational coupling constant, which was
previously set to 1. Thus, for a non-zero cosmological constant
$\Lambda$, we have a smooth massless limit (unitary in $AdS_4$), while
for non-zero mass we have a smooth flat limit. The special point 
$m^2 = 6\Lambda$ corresponds to the boundary of the unitarity region.
As in the massless case, the only way to obtain a non-trivial result
is by rescaling of the coupling constant, leaving us with
non-minimal interactions.

\section{Conclusion}

In this work, we investigate the gravitational interactions of
massive spin 3 (the first bosonic case, which requires non-minimal
corrections), including its massless and partially massless limits. To
find a general structure for non-minimal interactions, we consider the
partially massless case, and then use these results to construct the
minimal gravitational vertex (i.e. containing both standard minimal
interactions and non-minimal ones with the minimum number of
derivatives) for the massive case. As in the massive spin 5/2 case,
our result has a non-singular massless limit for a non-zero
cosmological constant, and a non-singular flat limit for a non-zero
mass. However, at the specific point $m^2 = 6\Lambda$, corresponding
to the boundary of the unitary allowed region, standard minimal
interactions disappear, leaving only non-minimal ones.

\end{document}